# Unusual Sign Reversal of Field-like Spin-Orbit Torque in Pt/Ni/Py with an Ultrathin Ni Spacer

Zishuang Li[1], Wenqiang Wang[1], Kaiyuan Zhou[1], Xiang Zhan[1], Tiejun Zhou[2,*], and Ronghua Liu[1,*]

[1]National Laboratory of Solid State Microstructures, School of Physics and Collaborative Innovation Center of Advanced Microstructures, Nanjing University, Nanjing 210093, China

[2]Centre for Integrated Spintronic Devices, School of Electronics and Information, Hangzhou Dianzi University, Hangzhou 310018, China

* E-mail: rhliu@nju.edu.cn, tjzhou@hdu.edu.cn

**Abstract**

The magnetization manipulation by spin-orbit torques (SOTs) in nonmagnetic-metal (NM)/ferromagnet (FM) heterostructures has provided great opportunities for spin devices. Besides the conventional spin Hall effect (SHE) in heavy metals with strong spin-orbit coupling, the orbital currents have been proposed to be another promising approach to generate strong SOTs. Here, we systematically study the SOTs efficiency and its dependence on the FM thickness and different NM/FM interfaces in two prototypical Pt/Py and Ta/Py systems by inserting an ultrathin magnetic layer (0.4 nm thick ML = Co, Fe, Gd, and Ni). The dampinglike (DL) torque efficiency $\xi_{DL}$ is significantly enhanced by inserting ultrathin Co, Fe, and Ni layers and is noticeably suppressed for the Gd insertion. Moreover, the Ni insertion results in a sign change of the field-like (FL) torque in Pt/Py and substantially reduces $\xi_{DL}$ in Ta/Py. These results are likely related to the additional spin currents generated by combining the orbital Hall effect (OHE) in the NM and orbital-to-spin conversion in the ML insertion layer and/or their interfaces, especially for the Ni insertion. Our results demonstrate that inserting ultrathin ML can effectively manipulate the strength and sign of the SOTs, which would be helpful for spintronics applications.

# I. INTRODUCTION

Current-induced SOTs in NM/FM heterostructures have emerged as a promising approach for electric control of magnetization reversal to achieve energy-efficient SOT-magnetoresistive random access memory (SOT-MRAM) and compensation of the intrinsic magnetic damping for self-sustained nano-oscillator [1-10]. The foundation of these SOTs in bilayers is commonly associated with the SHE in the NM layer with strong spin-orbit coupling (iSOC) [11] and/or interfacial Rashba-Edelstein effect (iREE) [12,13] at the NM/FM interface with an enhanced interfacial SOC (ISOC) due to the inversion symmetry breaking. Under an in-plane electric field along the longitudinal direction of NM/FM bilayers, an out-of-plane spin current with transverse spin polarization and a net spin accumulation near the interface can be generated by SHE and iREE. These generated spin currents flow into the FM layer through the interface, consequently applying torques on magnetization via the spin-transfer mechanism. The net spin polarization due to the interfacial spin accumulation exerts torques on the adjacent magnetization via the exchange interaction. Notably, recent theoretical and experimental works have shown that the orbital current, the flow of orbital angular momentum perpendicular to the charge current, can be generated by an electric field in a broader range of materials even without the requisition of SOC. Terming the orbital Hall effect [14,15], the currents differ from their spin counterparts as they cannot couple directly to the magnetization. Before exerting torque on magnetization, the orbital currents must be converted into spin currents via SOC in the NM, FM, and/or interface.

Based on torque and magnetization orientations, these spin-currents-generated SOTs are distinctly classified as damping-like (DL) torque [$\tau_{DL} \propto \mathbf{m} \times (\mathbf{m} \times \boldsymbol{\sigma})$] and field-like (FL) torque [$\tau_{FL} \propto (\mathbf{m} \times \boldsymbol{\sigma})$] [16,17]. In principle, the DL and FL torques arise from the absorption of the spin current component transverse to magnetization $\mathbf{m}$ and the reflection of spin current with some spin rotation, respectively. In addition, besides the spin currents generation mechanisms mentioned above, the high interfacial transparency is also vital to achieving efficient magnetization manipulation for the development of low-energy consumption spintronic devices because the effective DL torque efficiency is given by $\xi_{DL} = T_{int}\theta_{sh}$, where $T_{int}$ signifies the interfacial spin transparency and $\theta_{sh} = (2e/\hbar)j_s/j_c$ denotes the charge-to-spin current conversion efficiency, known as the spin Hall angle. Employing a simplified drift-diffusion analysis [16,17], the spin current diffuses into the adjacent FM layer via the mediation of electrons and diminishes sharply near the NM/FM interface due to spin backflow and spin memory loss [18] ($T_{int} = T_{int}^{SBF} \times T_{int}^{SML}$). According to the SOT-generation mechanisms above, it is a feasible scheme to achieve a high $\theta_{sh}$ and/or a large $T_{int}$ for a highly efficient manipulation of magnetization in practical spin devices.

Previously, much effort has been contributed to exploring high SOT efficiency through exploring spin-source materials with high intrinsic bulk $\theta_{sh}$ and tailoring the interface with enhanced spin transmissivity and substantial interfacial SOTs [10,19,20]. Besides the extensively studied heavy metals (HMs) with strong SOC, such as Pt [18,21], Ta [2,5,22], and W [23,24], other materials, including the topological insulator $Bi_xSe_{1-x}$ [25], $Bi_{1-x}Sb_x$ [26], and the highly conductive $Pt_{1-x}Au_x$ [27] alloys have been proved to exhibit high $\theta_{sh}$ or low-power consumption. On the other hand, interface engineering, e.g., insertion of ultrathin nonmagnetic layer like Hf [28], Mo [29], Cu [23,30,31], or two-dimensional van der Waals material $MoTe_2$ [32] between the HM and FM layers, oxygen-induced interface orbital hybridization [5], interfacial $H^+$ and $O^{2-}$ ion manipulations [33], utilization of antiferromagnetic (or paramagnetic) insulating NiO or $CoO_x$ thin layer [34-37], and magnetic metal spacer layer [38,39], has also been explored intensely. Several different mechanisms exist for these interface engineering methods (e.g., additional interface-generated SOTs, enhancing interfacial spin transparency, and orbital-to-spin current conversion) to enhance SOT efficiency. The experimentally quantitative measurement of the DL and FL torque efficiencies and identification of the underlying mechanisms in two prototypical Pt- and Ta-based systems are crucial for practical spintronics applications.

In this study, we present an investigation into the SOT efficiency dependence of the prototypical Pt/Py on different Pt/FM interfaces via inserting an ultrathin magnetic layer (ML= 0.4-nm-thickness Co, Fe, Gd, and Ni) sandwiched between Pt and Py layers and Ta/Ni(0.5 nm)/Py, using spin-torque ferromagnetic resonance (ST-FMR) technique with varying thickness of Py. Notably, the damping-like torque efficiency $\xi_{DL}$ is enhanced by 21% with Co, 43% with Fe, and 33% with Ni insertion layers and is suppressed by 17% with the Gd insertion layer. In contrast to the Pt/Py with a negligible field-like torque $\xi_{FL} \sim -0.002$, the insertion of an ultrathin magnetic layer results in a considerable $\xi_{FL} \sim -0.009 - -0.023$ for Gd, Fe, and Co, and its sign reversal for the Ni insertion system ($\xi_{FL} \sim 0.011$). For Ta/Ni/Py system, both $\xi_{FL}$ and $\xi_{DL}$ are significantly suppressed by inserting a 0.4-nm-thick Ni spacer but do not change their signs. Beyond the conventional SHE-generated spin currents observed in the HM, our results suggest the potential existence of spin currents originating from the conversion of orbital currents to spin currents via the OHE within the NM layer and SOC within the FM layer, especially for the Ni insertion systems.

## II. EXPERIMENTAL DETAILS

Multilayer films with the structure of Pt(6)/Py(*t*), Pt(6)/ML (0.4)/Py(*t*) (ML = Co, Fe, Gd, and Ni) (number in parentheses are thickness in nanometers) are deposited on annealed $Al_2O_3$ substrate with (0001) orientation by d.c. magnetron sputtering at room temperature. A 2-nm-thick MgO covers all multilayers to prevent oxidization in the air. To determine the SOT efficiency from the total ferromagnetic-layer thickness ($t_{FM}^{tot}$) dependence of the ST-FMR signal, we vary the Py thickness *t* from 3 nm to 8 nm for the Pt/Py, Pt/Co/Py, Pt/Fe/Py, and Pt/Gd/Py and from 3 nm to 10 nm for the Pt/Ni/Py. All the samples are patterned into 5 μm × 8 μm stripes with two top electrodes of Au (80 nm) for ST-FMR measurement, as shown in Fig. 1(a). For the ST-FMR measurement, an in-plane external field *H* is applied at an angle of $\varphi = 30°$ and $\varphi = 210°$ from the longitudinal direction of the stripes. To better eliminate the influence of experimental errors on the effective demagnetization field $4\pi M_{eff}$ and linewidth $\Delta H$, we used the average value of two directions to analyze all the data.

## III. RESULTS AND DISCUSSION

### A. Spin-torque efficiencies in Pt-based trilayers

Figs.1 (b) and (c) show the representative ST-FMR spectra of Pt(6)/Co(0.4)/Py(3) and Pt(6)/Ni(0.4)/Py(3) samples with excitation frequencies from 4 to 12 GHz and 6 to 14 GHz, respectively, at $\varphi = 30°$. The obtained $V_{\text{mix}}$ signal is fitted using a Lorentzian function [1,4,7]:

$$V_{\text{mix}} = V_S \frac{\Delta H^2}{[(H-H_{\text{res}})^2 + \Delta H^2]} + V_A \frac{\Delta H(H-H_{\text{res}})}{[(H-H_{\text{res}})^2 + \Delta H^2]} \quad (1)$$

Where $V_S$, $V_A$, $\Delta H$ and $H_{\text{res}}$ are the symmetric and antisymmetric Lorentzian components, the linewidth, and the resonance field, respectively. Based on the generated spin current with conventional spin polarization $\sigma_y$ via SHE/iREE, $V_S$ is proportional to the out-of-plane DL torque effective field $H_{\text{DL}}$, while $V_A$ is related to the sum of the in-plane Oersted field $H_{\text{Oe}}$ and in-plane FL torque effective field $H_{\text{FL}}$.

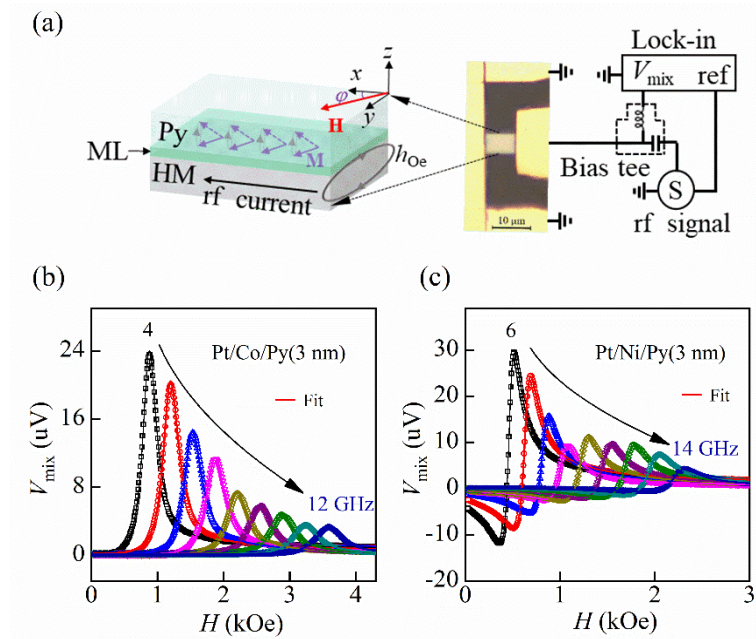

Figure 1. (a) Left: Illustration of the stack structure of multilayer, coordinate system, and magnetization dynamics in the ST-FMR measurement. Right: schematic diagram of the ST-FMR setup. (b) - (c) The representative ST-FMR spectra $V_{\text{mix}}$ obtained with the excitation frequencies from $f$ = 4 to 12 GHz and $f$ = 6 to 14 GHz with a step of 1 GHz at $\varphi$ = 30° for Pt(6)/Co(0.4)/Py(3) (b) and Pt(6)/Ni(0.4)/Py(3) (c) samples, respectively. The solid lines are the fitting results with Eq. (1).

To investigate the interface-related SOTs, we measured the ST-FMR spectra with various Py layer thicknesses (*t*) for the Pt/ML/Py trilayers. Figures 2(a)-(e) show the representative ST-FMR spectra $V_{\text{mix}}$ for Pt/Py, Pt/Co/Py, Pt/Fe/Py, Pt/Gd/Py, and (e) Pt/Ni/Py at $f$ = 9 GHz and $\varphi$ = 30°. The ratio of the magnitude of the symmetric and antisymmetric Lorentzian components ($V_S/V_A$) is extracted by fitting $V_{\text{mix}}$ spectra using Eq. (1). The FMR spin-torque generation efficiency $\xi_{\text{FMR}}$ is determined by the ratio of $V_S/V_A$ as follows [40-42]

$$\xi_{\text{FMR}} = \frac{V_S}{V_A} \frac{e\mu_0 M_s t_{\text{FM}}^{\text{tot}} t_{\text{HM}}}{\hbar} \sqrt{1 + 4\pi M_{\text{eff}}/H_{\text{res}}} \qquad (2)$$

, where $t_{\text{HM}}$, $t_{\text{FM}}^{\text{tot}}$, $e$, and $\hbar$ represent the thickness of the HM layer and the total FM layer, the electronic charge, and reduced Planck's constant, respectively. The effective demagnetization $4\pi M_{\text{eff}}$ is obtained by fitting the experimental dispersion of $f$ vs. $H_{\text{res}}$ using the Kittel formula[7,41]: $f = (\gamma/2\pi)\sqrt{H_{\text{res}}(H_{\text{res}} + 4\pi M_{\text{eff}})}$, where $\gamma/2\pi$ is the gyromagnetic ratio.

Figs. 2 (f) and (g) show the obtained $V_S/V_A$ and $\xi_{\text{FMR}}$ for all Pt/ML/Py samples with different Py layer thicknesses. For all these Pt-based samples, the sign of $V_S$ and $V_A$ is positive and keeps unchanged for all studied Py layer thicknesses, consistent with the positive spin Hall angle of Pt[18,27] and the dominant Oersted field torque against FL torque ($V_A \propto \tau_{\text{Oe}} - \tau_{\text{FL}}$) [Inset of Fig. 2(f)][43,44], or a superposition of the Oersted field and FL torques ($V_A \propto \tau_{\text{Oe}} + \tau_{\text{FL}}$) [Inset of Fig. 2(g)]. Compared to the Pt/Py without inserting layer, the ratio of $V_S/V_A$ exhibits a significant enhancement for the inserting Co, Fe, and Gd, but suppression for the inserting Ni. As shown in Fig. 2 (g), the $\xi_{\text{FMR}}$ can also be substantially tuned by inserting ultrathin ML metals. The ST-FMR efficiency $\xi_{\text{FMR}}$ is related to DL- and FL-SOT efficiencies [41]: $\xi_{\text{DL(FL)}} = (2e/\hbar)\mu_0 M_s t_{\text{FM}}^{\text{tot}} H_{\text{DL(FL)}}/j_c$, where $j_c$ represents the rf current density in the HM layer.

$\xi_{FMR}$ depends on $t_{FM}^{tot}$ because $H_{FL}$ is inversely proportional to the total FM layer thickness in terms of $H_{FL} \propto \xi_{FL}/t_{FM}^{tot}$ and $H_{Oe}$ is independent of $t_{FM}^{tot}$ if $j_c$ in the HM layer keeps the same [41,42]

$$\frac{1}{\xi_{FMR}} = \frac{1}{\xi_{DL}}(1 + \frac{\hbar}{e}\frac{\xi_{FL}}{\mu_0 M_s t_{FM}^{tot} t_{HM}}) \quad (3)$$

To determine the DL torque and FL torque efficiencies ($\xi_{DL}$ and $\xi_{FL}$), we plot the $1/\xi_{FMR}$ as a function of $1/t_{FM}^{tot}$ for all Pt-based samples in Fig.2 (h). Fig.2(h) shows that the sign of the intercept of all linear relations is positive, indicating that $\xi_{DL} > 0$ for all Pt-based samples, consistent with the positive spin Hall angle of Pt. In addition, the Pt/Co, Fe, Ni/Py samples show a smaller intercept value than the Pt/Py bilayer, indicating that the ultrathin ML (Co, Fe, Ni) insertion can significantly enhance $\xi_{DL}$. In contrast, the $\xi_{DL}$ of Pt/Gd/Py sample is suppressed, consistent with the previously reported Co(2)/Gd(t)/Pt(5) system [45]. The reason for $\xi_{DL}$ reduction is related to the fact that the primary SHE-generated and the secondary OHE-converted spin currents have opposite signs due to the positive spin and orbital Hall conductivities in Pt and negative SOC in Gd.

Furthermore, Fig. 2 (h) shows that the Pt/Ni/Py exhibits a positive slope which is opposite to the other four Pt-based systems, indicating that the sign of $\xi_{FL}$ is reversed by inserting an ultrathin Ni layer. We note that according to the linear dependence of $1/\xi_{FMR}$ on $1/t_{FM}^{tot}$ and Eq. (3), one can exclude the additional contributions, including the spin-pumping and thermal effects for $\xi_{DL}$ and $\xi_{FL}$. The determined values of $\xi_{DL}$ and $\xi_{FL}$ for all five series samples from Fig .2(h) are summarized in Table 1.

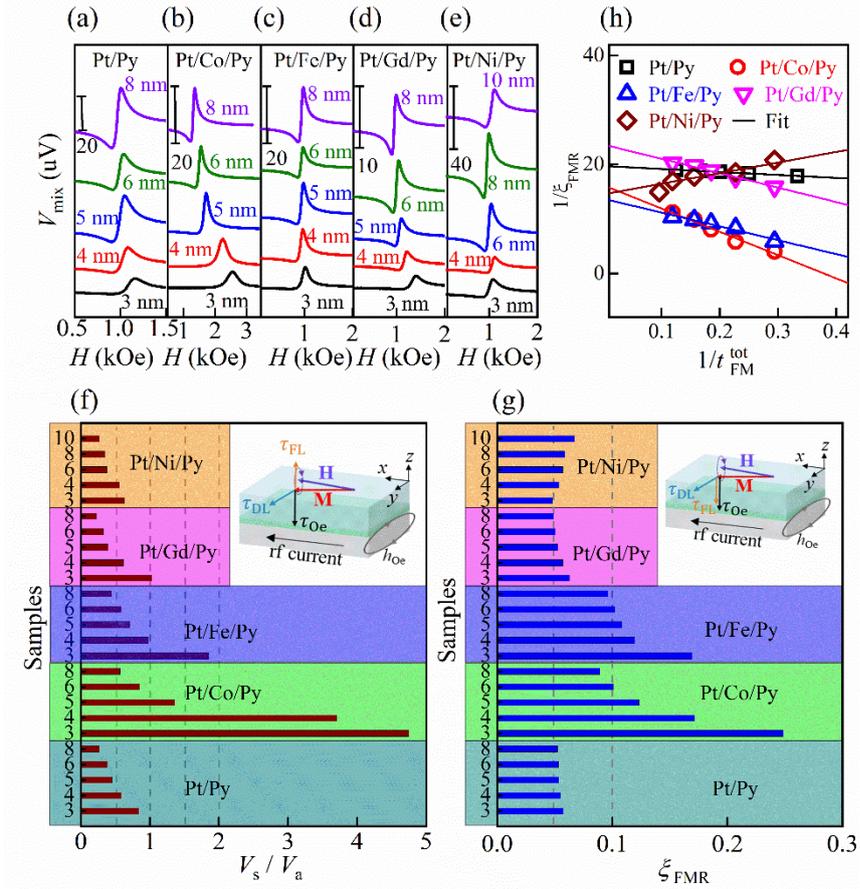

Figure 2. (a) - (e) The representative ST-FMR spectra of Pt/Py($t$) (a), Pt/Co/Py($t$) (b), Pt/Fe/Py($t$) (c), Pt/Gd/Py($t$) (d), and Pt/Ni/Py($t$) (e) at $f$ = 9 GHz and $\varphi = 30°$. The Py thickness $t$ is labeled in the panel. (f) - (g) The ratio of $V_S/V_A$ (f) and FMR-SOT efficiencies $\xi_{FMR}$ (g) for five different series of samples with various Py thicknesses. The insets illustrate the SOTs and Oersted field torque with the opposite (f) and same (g) direction to the FL torque. (h) The inverse of SOT efficiency ($1/\xi_{FMR}$) as a function of the inverse of the total magnetic layers thickness ($1/t_{FM}^{tot}$) for all Pt-based devices. The symbols and solid lines are the experimental data and the linear fitting results, respectively.

**B. Interfacial spin transport parameters in Pt-based trilayers**

According to the SOT-generation mechanism, the interface-generated spin currents and the interfacial spin transmission $T_{int}$ can significantly contribute to SOT efficiency besides the bulk SHE inside the Pt layer. To shed light on how the inserting ultrathin ML spacer impacts the DL and FL torque efficiencies, we further characterize the interfacial properties of the Pt/ML/Py trilayers by quantifying the interfacial magnetic anisotropy energy density $K_s$, the real and imaginary part of the effective spin-mixing conductance $\text{Re}(G_{\text{eff}}^{\uparrow\downarrow}), \text{Im}(G_{\text{eff}}^{\uparrow\downarrow})$, and the two-magnon scattering coefficient $\beta_{\text{TMS}}$. Firstly, the $K_s$ can be quantified from the $1/t_{\text{FM}}^{\text{tot}}$ dependence of $4\pi M_{\text{eff}}$ using the following formula [40,41]

$$4\pi M_{\text{eff}} = 4\pi M_s - \frac{2K_s}{\mu_0 M_s}\frac{1}{t_{\text{FM}}^{\text{tot}}} \tag{4}$$

We determined $4\pi M_{\text{eff}}$ by fitting the measured $f$ vs. $H_{\text{res}}$ using the Kittel formula, as shown in the representative Pt/Co/Py samples in Fig. 3 (a). Since the ML insertion layers are strongly exchange coupled to the adjacent Py layer, the magnetic layer ML/Py is considered as a single magnetic unit. For instance, the total saturation magnetization $M_s$ of the ML/Py is determined by the formula of $M_s = (M_s^{\text{Py}} t_{\text{Py}} + M_s^{\text{ML}} t_{\text{ML}})/(t_{\text{Py}} + t_{\text{ML}})$, where the $M_s^{\text{Py}}$ and $M_s^{\text{ML}}$ are the saturation magnetization of Py and ML, respectively. For the representative 6 nm-thickness samples, $M_s^{\text{Py}} = 716$ emu/cm³, $M_s^{\text{Co}} = 1130$ emu/cm³, $M_s^{\text{Fe}} = 1440$ emu/cm³, and $M_s^{\text{Ni}} = 382$ emu/cm³ are measured by the vibrating sample magnetometry (VSM), consistent with previously reported values [5,40,41]. The obtained $K_s$ by fitting the experimental results using Eq. (4) in Fig.3 (b) is summarized in Table 1. $K_s$ of Pt/Co/Py and Pt/Gd/Py is several times larger than that of Pt/Py, which is related to the strong interfacial PMA for Pt/Co due to the strong iSOC and interlayer antiferromagnetic coupling between Py and Gd [46-48]. However, for the Ni and Fe inserting layers, they have a small $K_s$ comparable to the Pt/Py sample.

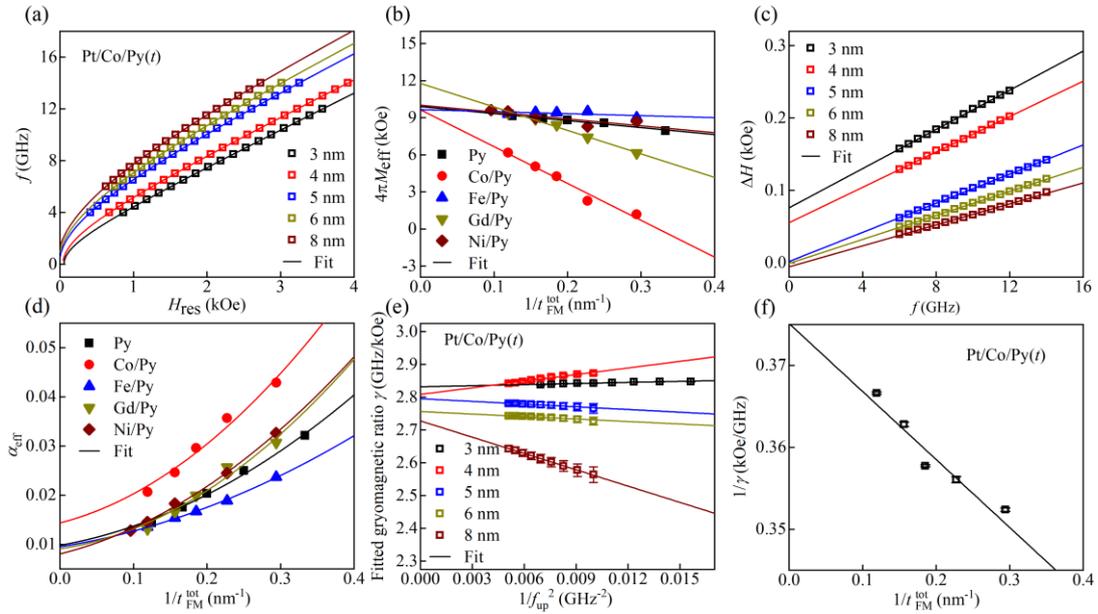

Figure 3. (a) The dispersion relation curves between $f$ and $H_{res}$ (symbols) and the corresponding Kittel fittings (solid curves) for the Pt/Co/Py($t$) samples. (b) The effective demagnetization field $4\pi M_{eff}$ as a function of $1/t_{FM}^{tot}$ for all five series of Pt/ML/Py($t$) samples. The solid lines are the linear fitting result. (c) Linewidth $\Delta H$ versus resonance frequency $f$. The solid lines are the linear fitting. (d) Damping $\alpha_{eff}$ as a function of $1/t_{FM}^{tot}$ (symbols). The lines are the fitting results using Eq.(5). (e) The fitted value of the gyromagnetic ratio $\gamma_{fit}$ as a function of $1/f_{up}^2$ for Pt/Co/Py($t$) samples. The lower fitting frequencies are fixed at 4 GHz and the error bars are also plotted. The solid lines are the linear fitting. (f) The inverse gyromagnetic ratio $1/\gamma$ as a function of the inverse FM thickness $1/t_{FM}^{tot}$. The solid line is the linear fitting.

Furthermore, the $\text{Re}(G_{eff}^{\uparrow\downarrow})$ and interfacial $\beta_{TMS}$ can also be characterized by analyzing the thickness-dependent magnetic damping. Fig.3(c) shows the linewidth $\Delta H$ of the ST-FMR spectra as a function for the representative Pt/Co/Py($t$) samples. The effective Gilbert damping $\alpha_{eff}$ can be determined by fitting these experimentally obtained linewidth $\Delta H$ data using the formula of $\Delta H = (2\pi/\gamma)\alpha_{eff} f + \Delta H_0$, where $\Delta H_0$ is the inhomogeneous linewidth. The effective Gilbert damping $\alpha_{eff}$ is further expressed as [49,50]

$$\alpha_{eff} = \alpha_{int} + \text{Re}(G_{eff}^{\uparrow\downarrow})\frac{g\mu_B h}{4\pi M_s e^2}\frac{1}{t_{FM}^{tot}} + \beta_{TMS}\frac{1}{t_{FM}^{tot2}}, \qquad (5)$$

where $g$ is the Lande factor, $\mu_B$ is the Bohr magnetron, and $h$ is Planck's constant. The first term is the total magnetic layer thickness-independent intrinsic Gilbert damping $\alpha_{int}$, the second term is related to the spin current loss via spin pumping into the Pt layer and being absorbed due to spin memory loss (SML) at the Pt/ML and ML/Py interfaces, and the third term is the contribution from the TMS process. The interface $\beta_{\text{TMS}}$ depends on $(2K_s/M_s)^2$ and/or the density of magnetic defect at the interfaces. To determine the $\text{Re}(G_{\text{eff}}^{\uparrow\downarrow})$ and interface $\beta_{\text{TMS}}$, we plot the obtained $\alpha_{\text{eff}}$ as a function of $1/t_{\text{FM}}^{\text{tot}}$ in Fig. 3(d). The values extracted by fitting $1/t_{\text{FM}}^{\text{tot}}$ dependence of $\alpha_{\text{eff}}$ using Eq. (5) are summarized in Table 1. Consistent with the large $K_S$ above, the intrinsic Gilbert damping for the Pt/Co/Py sample is more than 1.4 times that of the other four series. Meanwhile, the Pt/Co/Py has a larger ST-FMR linewidth obtained at $f$ = 6 GHz than other systems with $t$ = 6 nm.

Additionally, in FM/NM systems, the gyromagnetic ratio $\gamma$ is also modified by the spin current pumped by FMR [51,52].

$$\frac{1}{\gamma} = \frac{1}{\gamma_0}\left\{1 - \text{Im}(G_{\text{eff}}^{\uparrow\downarrow})\frac{g\mu_B h}{4\pi M_s e^2}\frac{1}{t_{\text{FM}}^{\text{tot}}}\right\}, \tag{6}$$

where $\gamma_0$ represents the intrinsic gyromagnetic ratio of the isolated FM. Thus, we can calculate the imaginary part by the linear fitting of the inverse gyromagnetic ratio on the inverse FM thickness. Prerequisitely, we have to determine the gyromagnetic ratio of samples with different FM thicknesses. Here, we follow the methodology presented in Ref-[53] and adopt an asymptotic analysis to the data obtained over a finite range of frequencies. The fitted value $\gamma_{\text{fit}}$ has a strong dependence on the range of the fitting frequency. With an identical lower bound on the data used in the fitting, the precise determination of $\gamma$ is the asymptotic value as the upper fitting frequency $f_{\text{up}} \to \infty$. Empirically, the $\gamma_{\text{fit}}$ is linearly proportional to $1/f_{\text{up}}^2$ and thus, the precise $\gamma$ is determined by the intercept of the linear fitting. Finally, we can obtain the imaginary part of the spin mixing conductance from the slope of the linear fitting of $1/\gamma$ on $1/t_{\text{FM}}^2$. The calculating results of Pt/Co/Py($t$) are shown in Fig.3(e)(f) and the other results are also summarized in Table 1.

From the DL and FL torque efficiencies and the interface-dependent spin-transport parameters in Table 1, we deduce that the enhanced SOTs efficiencies with Co insertion are the result of competition between the enhancement of iREE-generated SOTs and $\beta_{\text{TMS}}$. Both of these are closely related to the strong iSOC at the Pt/Co interface. For the Fe insertion, the smaller $\beta_{\text{TMS}}$ and negligible $K_s$ indicate that the Pt/Fe interface can improve the interfacial spin transparency and/or the interfacial spin accumulation via enhancing the exchange interaction for more considerable SOTs efficiencies. Although the enhanced iSOC may improve the interface-generated SOTs for the Gd insertion, the observed suppression of the DL torque could be related to the poor interfacial spin transparency due to the robust AFM coupling between Py and Gd. Besides, according to previous experiments and theoretical calculations [45, 51], it is noteworthy that Pt exhibits both positive spin Hall and orbital Hall ratios, while Gd has a negative spin-orbit coupling ($\langle \boldsymbol{L} \cdot \boldsymbol{S} \rangle < 0$). The orbital current generated in the Pt sublayer is converted into spin currents with spin polarization, opposite to the conventional SHE-generated spin currents, by the Gd spacer and thus suppresses the total DL torque efficiency.

In the Pt/Ni/Py configuration, the enhancement of $\xi_{\text{DL}}$ may be linked to the improvement of the interfacial spin transparency and the additional interface-generated spin currents, e.g., orbital current and orbital-to-spin conversion in the Pt/Ni and Ni sublayer [54,55]. In contrast, the positive $\xi_{\text{FL}}$ is opposite to other four Pt-based systems. Unlike the DL torque related to the absorption of spin current, the FL torque arises from the reflection of the spin current at the HM/FM interfaces. Therefore, the FL torque is more sensitive to the change in the interfacial electronic structure by inserting different ML spacers. According to previously reported sign reversal of SOTs due to the orbital torque in the Nb/Ni, Ta/Ni, and Cr/Ni bilayers [54-56], in our case, the sign reversal of $\xi_{\text{FL}}$ observed in the ultrathin Ni insertion system would also be correlated to the secondary spin currents due to orbital-to-spin conversion occurring at the Pt/Ni interface or inside the Ni spacer, since Ni possesses the largest orbital-to-spin conversion efficiency compared with the other four ML according to the first-principle calculated results [54].

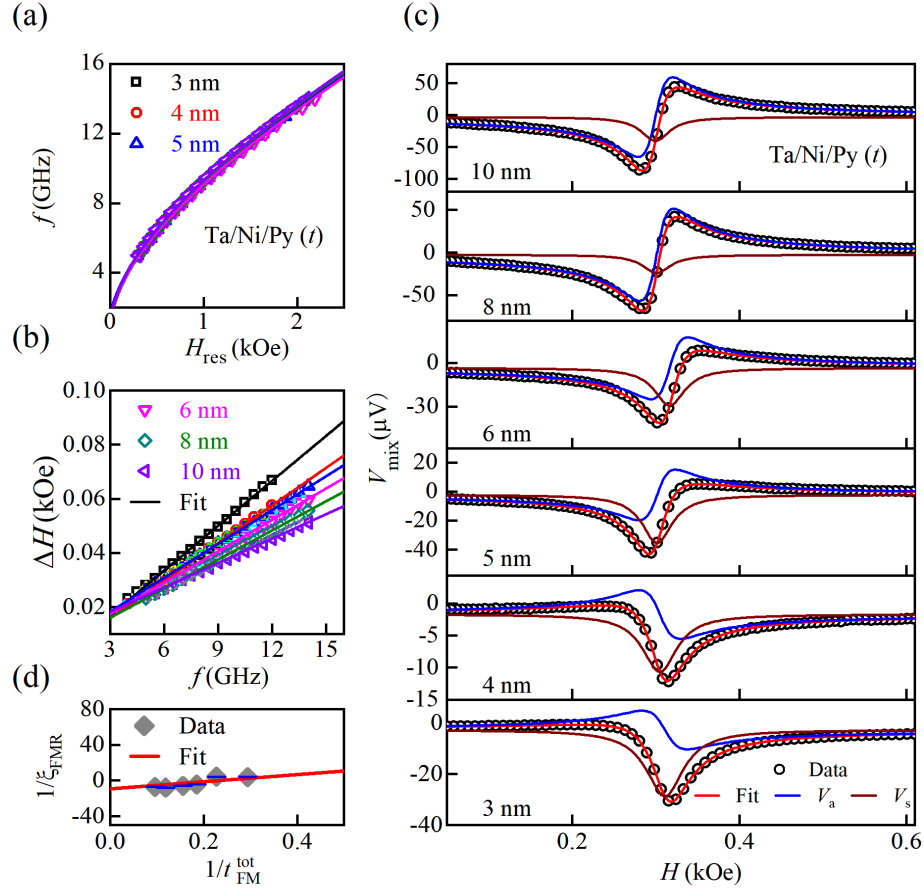

Figure 4. (a) The experimental dispersion curves of $f$ vs $H_{\text{res}}$ (symbols) and the corresponding Kittel fittings (solid curves) for a series of Ta(6)/Ni(0.4)/Py($t$) samples with the labeled Py thickness $t$. (b) Linewidth $\Delta H$ versus resonance frequency $f$. The solid lines are the linear fittings. (c) The typical ST-FMR spectra $V_{\text{mix}}$ obtained at $f = 5$ GHz and the corresponding Lorentzian fitting curves using Eq. (1). (d) The experimental data of $1/\xi_{\text{FMR}}$ vs. $1/t_{\text{FM}}^{\text{tot}}$, and its linear fitting using Eq. (5). The error bars are significantly smaller than the data symbols.

## C. Spin-orbit torques in Ta/Ni/Py trilayers

To further investigate the sign reversal of field-like torque caused by inserting an ultrathin Ni spacer in the Pt/Py system, we adopt the Ta(6)/Ni(0.4)/Py($t$) system as a control experiment. Figure 4(a) shows that the corresponding six dispersion curves of a series of Ta/Ni/Py($t$) with $t$ = 3, 4, 5, 6, 8, and 10 nm samples collapse into a single curve, which the Kittel formula can fit well with nearly the same parameter $M_{\text{eff}} = 780$ emu/cm$^3$, suggesting a good consistency for the series of samples and a weak interfacial PMA energy density $K_s$ at Ta/Ni interface. The effective Gilbert damping $\alpha_{\text{eff}}$ can be determined by linearly fitting the linewidth $\Delta H$ vs. $f$ data in Fig.4(b). Furthermore, $\alpha_{\text{int}}$, $\beta_{\text{TMS}}$, $\text{Re}(G_{\text{eff}}^{\uparrow\downarrow})$ and $\text{Im}(G_{\text{eff}}^{\uparrow\downarrow})$ can be extracted by fitting $1/t_{\text{FM}}^{\text{tot}}$ dependence of $\alpha_{\text{eff}}$ and $1/\gamma$ using Eq. (5)(6), respectively, as the same as in Fig. 3(d)(f). These values for Ta/Ni/Py trilayer system are also summarized in Table 1.

Figure 4(c) shows the representative ST-FMR spectra obtained at $f$ = 5 GHz and their Lorentzian fitting curves with Eq. (1). The blue and brownish red represent the antisymmetric $V_A$ and symmetric components $V_S$ of the $V_{\text{mix}}$ spectra, respectively. In contrast to the Pt/Ni/Py system, the $V_s$ signal reflecting the DL torque is negative and opposite to the Pt-based system in Fig. 2(e), consistent with the expectation because of the negative spin Hall angle for Ta. The negative intercept $\xi_{DL} = -0.108$ of the linear fitting $1/\xi_{\text{FMR}}$ vs. $1/t_{\text{FM}}^{\text{tot}}$ using Eq. (3), as shown in Fig. 4(d), also confirms it. More surprisingly, we find the $V_A$ signal reversal during increasing Py thickness from $t \leq 4$ to $t \geq 5$ nm. Based on the discussion of Pt-based systems above, the Oersted field torque $\tau_{\text{Oe}}$ has the dominant contribution to $V_A$ for all studied Py thicknesses samples because Pt has a higher conductance than the studied ML. However, for Ta/Ni/Py system, the Oersted field $H_{\text{Oe}}$ is expected to be reduced significantly if the total $rf$ input current keeps the same for Pt- and Ta-based systems because Ta has more than ten times lower conductance than Pt. In addition, the FL torque is inversely proportional to the thickness of the FM layer $H_{\text{FL}} = \frac{\hbar}{2e} \frac{\xi_{\text{FL}} J_{\text{rf}}}{4\pi M_s t_{\text{FM}}}$, thus decreases with increasing Py thickness, while the current-induced Oersted field on the FM layer $H_{\text{Oe}} = \frac{J_{\text{rf}} t_{\text{NM}}}{2}$ is approximately linearly proportional to the thickness of the underlying NM layer and remains identical in the series with a constant thickness of the Ta layer [41]. Therefore, the $V_A$ signal reversal with $t_{\text{FM}}$ infers that the considerable FL torque $\tau_{\text{FL}}$ and $\tau_{\text{Oe}}$ have the opposite direction in the Ta/Ni/Py system [43,44], as illustrated in the inset of Fig. 2(f).

Next, from the linear fitting $1/\xi_{\text{FMR}}$ vs. $1/t_{\text{FM}}^{\text{tot}}$ using Eq. (3), we obtain the torque efficiencies $\xi_{\text{DL}} = -0.108$ and $\xi_{\text{FL}} = -0.034$ for Ta(6)/Ni(0.4)/Py(t) system. Contrary to the Pt/Ni/Py discussed above, both DL and FL torques in Ta/Ni/Py have the same sign as the pure Ta/Py. However, the previously reported Ta/Ni and Nb/Ni exhibit the positive sign for both DL and FL torques, opposite to Ta(6)/Ni(0.4)/Py(t), Ta/Py and Nb/Py [55]. The reported sign reversal of DL and FL torques in Ta/Ni and Nb/Ni bilayers indicates that a strong orbital current is generated in Ta and Nb sublayers and then converted into the spin currents inside the Ni sublayer [52]. Furthermore, previous studies report that the orbital-to-spin conversion length is generally several nanometers [57,58]. Therefore, in our case, the 0.4-nm thick Ni insertion layer is too thin to convert the input orbital current to spin current completely and sufficiently. Thus, the sign of $\xi_{FL}$ in Ta/Ni/Py remains unchanged. Additionally, the enhancement (suppression) of $\xi_{DL}$ in Pt/Ni/Py (Ta/Ni/Py) is also consistent with the expected positive sign of orbital torque in Pt- and Ta-based systems because, unlike spin Hall conductivity, the orbital Hall conductivity is positive for both Pt and Ta [45].

Table 1. The parameters of the DL torque efficiency $\xi_{DL}$, the FL torque efficiency $\xi_{FL}$, -the interfacial PMA energy density $K_s$, two-magnon scattering coefficient $\beta_{TMS}$, the real and imaginary part of the total effective spin-mixing conductance $\text{Re}(G_{\text{eff}}^{\uparrow\downarrow})$ and $\text{Im}(G_{\text{eff}}^{\uparrow\downarrow})$, and the intrinsic magnetic damping constant $\alpha_{\text{int}}$ for the Pt/Py, Pt/Co/Py, Pt/Fe/Py, Pt/Gd/Py, Pt/Ni/Py, and Ta/Ni/Py systems

| Sample | $\xi_{DL}$ | $\xi_{FL}$ | $K_s$ (erg/cm$^2$) | $\beta_{TMS}$ (nm$^2$) | $\text{Re}(G_{\text{eff}}^{\uparrow\downarrow})$ ($10^{15}\Omega^{-1}\text{m}^{-2}$) | $\text{Im}(G_{\text{eff}}^{\uparrow\downarrow})$ ($10^{15}\Omega^{-1}\text{m}^{-2}$) | $\alpha_{\text{int}}$ |
|---|---|---|---|---|---|---|---|
| Pt/Py | 0.051 | -0.002 | 0.255 | 0.124 | 0.500 | -1.461 | 0.010 |
| Pt/Co/Py | 0.062 | -0.023 | 1.402 | 0.211 | 0.747 | 0.455 | 0.014 |
| Pt/Fe/Py | 0.073 | -0.016 | 0.078 | 0.082 | 0.482 | -0.340 | 0.010 |
| Pt/Gd/Py | 0.042 | -0.009 | 0.856 | 0.185 | 0.415 | 0.104 | 0.009 |
| Pt/Ni/Py | 0.068 | 0.011 | 0.242 | 0.157 | 0.675 | -0.096 | 0.008 |
| Ta/Ni/Py | -0.108 | -0.034 | 0.060 | 0.052 | -0.223 | -1.194 | 0.008 |

**D. Discussion of sign reversal of $\xi_{FL}$ in Pt/Ni bilayer and Pt/Ni/Py trilayer**

There have been some preceding studies on the $\xi_{FL}$ in various systems, mainly focusing on the influence of interfacial effects. For example, in the Pt/Ni and Pt/Fe bilayer systems, the $\xi_{DL}$ is almost identical, while the $\xi_{FL}$ exhibits opposite signs despite the same underlying Pt sublayer [40]. In Ref-40, this is attributed to the sign change of the imaginary part of the spin-mixing conductance resulting from different interfacial electronic structures. We also plot all the calculated results of both $\xi_{DL}$, $\mathrm{Re}(G_{\mathrm{eff}}^{\uparrow\downarrow})$ and $\xi_{FL}$, $\mathrm{Im}(G_{\mathrm{eff}}^{\uparrow\downarrow})$ of all studied samples in Figure 5. According to the previous theories based on the drift-diffusion approximation [16,17,41], in the absence of spin memory loss, the torque efficiencies have a relation with the spin mixing conductance as

$$\xi_{DL} = \theta_{SH}\mathrm{Re}\left\{\frac{2G_{\mathrm{eff}}^{\uparrow\downarrow}\tanh\left(\frac{d_{NM}}{\lambda_{s,NM}}\right)}{G_{NM}\coth\left(\frac{d_{NM}}{\lambda_{s,NM}}\right)}\right\}, \xi_{FL} = \theta_{SH}\mathrm{Im}\left\{\frac{2G_{\mathrm{eff}}^{\uparrow\downarrow}\tanh\left(\frac{d_{NM}}{\lambda_{s,NM}}\right)}{G_{NM}\coth\left(\frac{d_{NM}}{\lambda_{s,NM}}\right)}\right\} \quad (7)$$

, where $G_{NM} \equiv \sigma_{NM}/\lambda_{s,NM}$ is the spin conductance of the NM. Even considering the potential strong spin scattering and spin memory loss induced by the ultrathin FM insertion layers, the torque efficiencies should retain a general proportionality to the spin mixing conductance. As depicted in Fig.5(a), the $\mathrm{Re}(G_{eff}^{\uparrow\downarrow})$ can roughly interpret the $\xi_{DL}$ variation induced by ultrathin FM insertion layer. However, the $\mathrm{Im}(G_{eff}^{\uparrow\downarrow})$ fails to reproduce the changes in $\xi_{FL}$, either the rise-and-fall trends or the positive-negative sign characteristics. Such a discrepancy highlights the possibility that our experimental results might go beyond the conventional interpretation, which primarily takes into account the spin degree, including either the SHE in the bulk NM or interface-generated spin currents, but neglects the orbital degree.

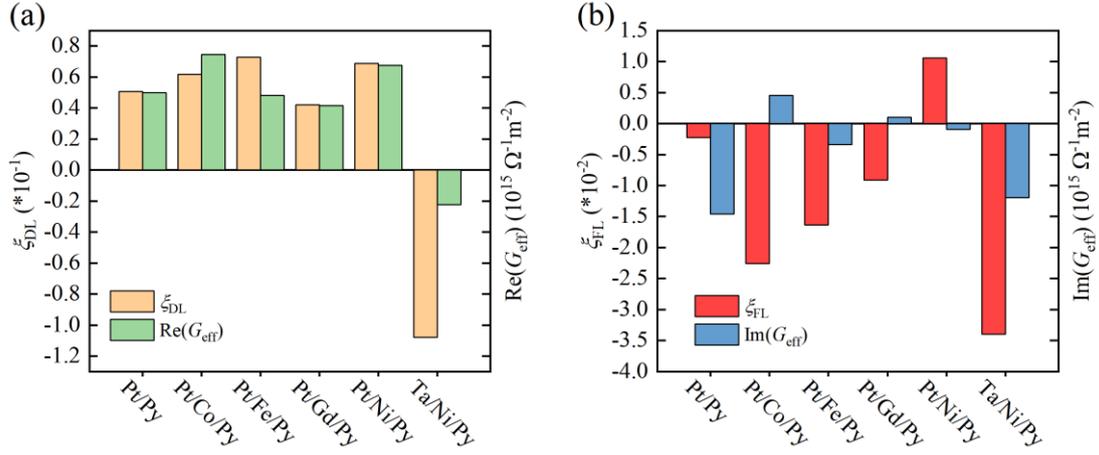

Figure 5. Calculated results of (a) damping like torque efficiency $\xi_{DL}$ and the real part of the effective spin mixing conductance $\mathrm{Re}(G_{\mathrm{eff}}^{\uparrow\downarrow})$, and (b) field like torque efficiency $\xi_{FL}$ and the imaginary part of the effective spin mixing conductance $\mathrm{Im}(G_{\mathrm{eff}}^{\uparrow\downarrow})$, respectively.

As for the Ta/Ni and Nb/Ni bilayers mentioned above [54,55], both $\xi_{DL}$ and $\xi_{FL}$ reverse their signs compared to Ta/Py and Nb/Py, as summarized in Table 2. The widely recognized reason is that the Ta/Ni and Nb/Ni systems generate considerable orbital currents-converted spin currents with the polarization opposite to the SHE-generated spin currents because of a high orbital-to-spin conversion efficiency in the Ni sublayer and a positive giant orbital Hall conductivity $\sigma_{OH}$ in the Ta and Nb sublayers [54,55], which differs from both positive signs for spin Hall and orbital Hall conductivities in the Pt. In our studied Ta/Ni/Py trilayer with a 0.4-nm-thick Ni spacer, the value of both $\xi_{DL}$ and $\xi_{FL}$ is reduced significantly compared to the Ta/Py bilayer, but their signs remain unchanged. The reason is that the ultrathin Ni spacer is thinner than the orbital-to-spin conversion length of several nanometers [57,58], so the orbital currents generated by the Ta sublayer are not fully converted into spin currents in only 0.4-nm-thick Ni spacer and cannot wholly surpass the SHE-generated spin currents with the negative sign and reverse the sign of $\xi_{DL}$ and $\xi_{FL}$ as the previously reported Ta/Ni bilayer [54]. Different from Ta/Ni/Py trilayer, our studied Pt/Ni/Py system exhibits a sign change of the $\xi_{FL}$, as the same as the previous Pt/Ni bilayer.

However, according to Hund's rules and first-principle calculated results [54], Fe, Co and Ni all exhibit positive orbital-to-spin conversion efficiency, indicating that the spin current converted from orbital currents by the Ni sublayer shares the same polarization with the SHE-generated spin currents within the Pt sublayer. Thus, it might be insufficient to account for the observed sign reversal of $\xi_{FL}$ within the previously reported Pt/Ni bilayer and our Pt/Ni/Py trilayer if only considering the signs of the spin Hall and orbital Hall conductivities of the Pt sublayer and the orbital-to-spin conversion of the Ni sublayer, as summarized in Table 3. Based on all sign results of $\xi_{DL}$ and $\xi_{FL}$ for various systems in Table 2, we propose a possible reason for the sign reversal of $\xi_{FL}$ and the enhancement of $\xi_{DL}$ in Pt/Ni/Py. This explanation considers the influence of orbital-to-spin conversion occurring within the Ni insertion layer or at the Pt/Ni interface, thereby contributing to $\xi_{FL}$ with a polarization opposite to that of SHE-generated spin currents within the Pt sublayer or other potentially underestimated origins impacting the FL torque. Different from the prominent $\xi_{FL}$ in Ta-based systems, Pt-based systems exhibit a significantly smaller $\xi_{FL}$ than $\xi_{DL}$ so that the emergence of secondary abnormal attributions to the FL torque can be easily observed. Our experimentally observed abnormal sign reversal of the FL torque in the Pt/Ni/Py system warrants further theoretical and modeling studies of spin Hall, orbital Hall effects, the conversion between them, and the relationship between them and spin torques in the most studied NM/FM bilayer systems.

Apart from the orbital torque mechanism, we also analyze other possible interface-related spin-dependent scattering mechanisms, particularly the spin-filtering effect [59, 60]. The spin filtering effect originates from the interfacial momentum-dependent spin-orbit field, such as the Rashba or Dresselhaus effective fields [12,13]. The electrons with spins parallel or antiparallel to these interfacial spin-orbit fields experience different scattering amplitudes. Consequently, the reflected and transmitted electrons become spin-polarized even if the incoming electrons are unpolarized. However, this scenario diverges from our experimental configuration, where the spin Hall effect in the underlying Pt layer generates the spin currents with a transverse spin polarization. These spin-polarized currents might substantially encounter distinct scattering amplitudes when passing through the Pt/ML and ML/Py interfaces, resulting in different spin transparencies, which have already been partly characterized through the spin mixing conductance above. It is also important to note that even if the spin filtering effect or interface-generated spin currents play an important role [16], the generated spin currents would naturally diffuse and flow into the adjacent FM layer and modulate both the damping-like and field-like torque efficiencies. Also note that the interfaces between Pt and Fe, Co, Ni exhibit inversion symmetry breaking and the same heavy metal Pt sublayer. These attributes give rise to the Rashba effective field with the same orientation due to the expected surface potential gradient with the same direction, implying that subsequent modifications in torque efficiencies have the same tendency across these interfaces [16,17, 60]. This perspective, however, is inconsistent with our experimental results, that only the field-like torque efficiency exhibits a sign change with Ni insertion alone.

Table 2. Signs of the damping-like torque efficiency $\xi_{DL}$ and the field-like torque efficiency $\xi_{FL}$ in selected systems.

|         | Pt/Fe | Pt/Fe(0.4)/Py | Pt/Ni | Pt/Ni(0.4)/Py | Ta/Py | Ta/Ni | Ta/Ni(0.4)/Py | Nb/Py | Nb/Ni |
|---------|-------|---------------|-------|---------------|-------|-------|---------------|-------|-------|
| $\xi_{DL}$ | +     | +             | +     | +             | −     | +     | −             | −     | +     |
| $\xi_{FL}$ | −     | −             | +     | +             | −     | +     | −             | −     | +     |
| Refs    | 40    | this          | 40    | this          | 51, 52| 51,52 | this          | 52    | 52    |

Table 3. Signs of orbital-to-spin conversion efficiency (spin-orbit coupling) $\langle \mathbf{L} \cdot \mathbf{S} \rangle$, spin Hall conductivity $\sigma_{SH}$ and orbital Hall conductivity $\sigma_{OH}$ of selected elements.

|                              | Fe | Co | Ni | Gd | Pt | Ta    | Nb |
|------------------------------|----|----|----|----|----|-------|----|
| $\langle \mathbf{L} \cdot \mathbf{S} \rangle$ | +  | +  | +  | −  | +  |       |    |
| $\sigma_{SH}$                | +  | +  | +  | −  | +  | −     | −  |
| $\sigma_{OH}$                |    |    | +  |    | +  | +     | +  |
| Refs                         | 51 | 45 | 45 | 45 | 51 | 51, 52| 52 |

## IV. CONCLUSION

In summary, our study presents a comprehensive examination of the SOT efficiencies in the HM/ML/Py trilayers with 0.4-nm-thick Co, Fe, Ni, and Gd spacer, employing the ST-FMR spectra technique. Compared to the Pt/Py bilayer with

$\xi_{DL} \sim 0.051$, we find that the DL torque efficiency $\xi_{DL}$ is significantly enhanced or suppressed by inserting an ultrathin ML spacer to modulate the interfacial spin transparency and/or the interface-generated spin currents, including iREE and orbital-to-spin conversion. Unlike the DL torque, the FL torque is significantly enhanced for all ML inserting systems and even changes its sign in the Ni insertion system. In contrast with Pt/Ni/Py system, Ta/Ni/Py system demonstrates a notable reduction in $\xi_{DL}$, but does not change the sign of SOTs by inserting a 0.4-nm-thick Ni spacer. The specific results for the ultrathin Ni insertion systems may find explanation in the secondary spin currents due to the OHE-generated orbital currents in the NM, subsequently converted to spin currents in the Ni insertion layer and/or Pt/Ni and Ta/Ni interfaces. Further theories and numerous modeling are essential to clarify these various spin current generation mechanisms, especially for orbital-to-spin current conversion and spin polarization directions in the FM and its interface, which would facilitate the electrical control of perpendicular nanomagnets for SOT-MRAM with promising high performance.

## Acknowledgments

The project is supported by the National Natural Science Foundation of China (Grants No. 12074178), the Open Research Fund of Jiangsu Provincial Key Laboratory for Nanotechnology, and the Key Research and Development Program of Zhejiang Province under Grant No. 2021C01039.

## References

[1] L. Q. Liu, T. Moriyama, D. C. Ralph, and R. A. Buhrman, Spin-torque ferromagnetic resonance induced by the spin Hall effect, Phys. Rev. Lett. **106**, 036601, (2011).
[2] L. Q. Liu, C. F. Pai, Y. Li, H. W. Tseng, D. C. Ralph, R. A. Buhrman , Spin-torque switching with the giant spin Hall effect of tantalum, Science **336**, 555, (2012).
[3] Q. Fu, L. Liang, W. Wang, L. Yang, K. Zhou, Z. Li, C. Yan, L. Li, H. Li, and R. Liu, Observation of nontrivial spin-orbit torque in single-layer ferromagnetic metals, Phys. Rev. B **105**, 224417, (2022).
[4] L. Yang, Y. Gu, L. Chen, K. Zhou, Q. Fu, W. Wang, L. Li, C. Yan, H. Li, L. Liang, Z. Li, Y. Pu, Y. W. Du, and R.H. Liu, Absence of spin transport in amorphous YIG evidenced by nonlocal spin transport experiments, Phys. Rev. B **104**, 144415, (2021).

[5] L. Yang, Y. Fei, K. Zhou, L. Chen, Q. Fu, L. Li, C. Yan, H. Li, Y. W. Du, and R.H. Liu, Maximizing spin–orbit torque efficiency of Ta(O)/Py via modulating oxygen-induced interface orbital hybridization, Appl. Phys. Lett. **118**, 032405, (2021).

[6] L. Li, L. Chen, R. H. Liu, and Y. Du, Recent progress on excitation and manipulation of spin-waves in spin Hall nano-oscillators, Chin. Phys. B **29**, 117102, (2020).

[7] W. Wang, Q. Fu, K. Zhou, L. Chen, L. Yang, Z. Li, Z. Tao, C. Yan, L. Liang, X. Zhan, Y. Du, and R. H. Liu, Unconventional spin currents generated by the spin-orbit precession effect in perpendicularly magnetized Co−Tb ferrimagnetic system, Phys. Rev. Appl. **17**, 034026, (2022).

[8] M. Wang, W. Cai, D. Zhu, Z. Wang, J. Kan, Z. Zhao, K. Cao, Z. Wang, Y. Zhang, T. Zhang, C. Park, J.-P. Wang, A. Fert, and W. Zhao, Field-free switching of a perpendicular magnetic tunnel junction through the interplay of spin–orbit and spin-transfer torques, Nat. Electron. **1**, 582, (2018).

[9] I. M. Miron, K. Garello, G. Gaudin, P. J. Zermatten, M. V. Costache, S. Auffret, S. Bandiera, B. Rodmacq, A. Schuhl, and P. Gambardella, Perpendicular switching of a single ferromagnetic layer induced by in-plane current injection, Nature **476**, 189, (2011).

[10] J. Ryu, S. Lee, K. J. Lee, and B. G. Park, Current-induced spin-orbit torques for spintronic applications, Adv. Mater. **32**, 1907148, (2020).

[11] J. E. Hirsch, Spin Hall effect, Phys. Rev. Lett. **83**, 1834, (1999).

[12] I. M. Miron, G. Gaudin, S. Auffret, B. Rodmacq, A. Schuhl, S. Pizzini, J. Vogel, and P. Gambardella, Current-driven spin torque induced by the Rashba effect in a ferromagnetic metal layer, Nat. Mater. **9**, 230, (2010).

[13] J. C. Sanchez, L. Vila, G. Desfonds, S. Gambarelli, J. P. Attane, J. M. De Teresa, C. Magen, and A. Fert, Spin-to-charge conversion using Rashba coupling at the interface between non-magnetic materials, Nat. Commun. **4**, 2944, (2013).

[14] D. Go, D. Jo, C. Kim, and H.-W. Lee, Intrinsic spin and orbital Hall effects from orbital texture, Phys. Rev. Lett. **121**, 086602, (2018).

[15] D. Jo, D. Go, and H.-W. Lee, Gigantic intrinsic orbital Hall effects in weakly spin-orbit coupled metals, Phys. Rev. B **98**, 214405, (2018).

[16] V. P. Amin and M. D. Stiles, Spin transport at interfaces with spin-orbit coupling: Phenomenology, Phys. Rev. B **94**, 104420, (2016).

[17] P. M. Haney, H.-W. Lee, K.-J. Lee, A. Manchon, and M. D. Stiles, Current induced torques and interfacial spin-orbit coupling: Semiclassical modeling, Phys. Rev. B **87**, 174411, (2013).

[18] L. Zhu, D. C. Ralph, and R. A. Buhrman, Maximizing spin-orbit torque generated by the spin Hall effect of Pt, Appl. Phys. Rev. **8**, 031308, (2021).

[19] Q. Shao, P. Li, L. Liu, H. Yang, S. Fukami, A. Razavi, H. Wu, K. Wang, F. Freimuth, Y. Mokrousov, M. D. Stiles, S. Emori, A. Hoffmann, J. Akerman, K. Roy, J.-P. Wang, S.-H. Yang, K. Garello, and W. Zhang, Roadmap of spin–orbit torques, IEEE Trans. Magn. **57**, 800439, (2021).

[20] X. Han, X. Wang, C. Wan, G. Yu, and X. Lv, Spin-orbit torques: Materials, physics, and devices, Appl. Phys. Lett. **118**, 120502, (2021).

[21] Y. Wang, P. Deorani, X. Qiu, J. H. Kwon, and H. Yang, Determination of intrinsic spin Hall angle in Pt, Appl. Phys. Lett. **105**, 152412, (2014).

[22] D. Velázquez Rodriguez, J. E. Gómez, L. Morbidel, P. A. Costanzo Caso, J. Milano, and A. Butera, High spin pumping efficiency in $Fe_{80}Co_{20}$/Ta bilayers, J. Phys. D-Appl. Phys. **54**, 325002, (2021).

[23] P. Yang, Q. Shao, G. Yu, C. He, K. Wong, X. Lu, J. Zhang, B. Liu, H. Meng, L. He, K. L. Wang, and Y. Xu, Enhancement of the spin–orbit torque efficiency in W/Cu/CoFeB heterostructures via

interface engineering, Appl. Phys. Lett. **117**, 082409, (2020).

[24] S. K. Li, X. T. Zhao, W. Liu, Y. H. Song, L. Liu, X. G. Zhao, and Z. D. Zhang, Interface effect of ultrathin W layer on spin-orbit torque in Ta/W/CoFeB multilayers, Appl. Phys. Lett. **114**, 082402, (2019).

[25] M. Dc, R. Grassi, J. Y. Chen, M. Jamali, D. Reifsnyder Hickey, D. Zhang, Z. Zhao, H. Li, P. Quarterman, Y. Lv, M. Li, A. Manchon, K. A. Mkhoyan, T. Low, and J. P. Wang, Room-temperature high spin-orbit torque due to quantum confinement in sputtered $Bi_xSe_{1-x}$ films, Nat. Mater. **17**, 800, (2018).

[26] N. H. D. Khang, Y. Ueda, and P. N. Hai, A conductive topological insulator with large spin Hall effect for ultralow power spin-orbit torque switching, Nat. Mater. **17**, 808, (2018).

[27] L. Zhu, D. C. Ralph, and R. A. Buhrman, Highly efficient spin-current generation by the spin Hall effect in $Au_{1-x}Pt_x$, Phys. Rev. Appl. **10**, 031001(R)(2018).

[28] C.-F. Pai, M.-H. Nguyen, C. Belvin, L. H. Vilela-Leão, D. C. Ralph, and R. A. Buhrman, Enhancement of perpendicular magnetic anisotropy and transmission of spin-Hall-effect-induced spin currents by a Hf spacer layer in W/Hf/CoFeB/MgO layer structures, Appl. Phys. Lett. **104**, 082407, (2014).

[29] H. K. Gweon, K.-J. Lee, and S. H. Lim, Spin-orbit torques and their angular dependence in ferromagnet/normal metal heterostructures, Appl. Phys. Lett. **115**, 122405, (2019).

[30] F. Liu, C. Zhou, R. Tang, G. Chai, and C. Jiang, Controllable charge-spin conversion by Rashba-Edelstein effect at Cu/Ta interface, J. Magn. Magn. Mater. **540**, 168462, (2021).

[31] L. Ni, Z. Chen, X. Lu, Y. Yan, L. Jin, J. Zhou, W. Yue, Z. Zhang, L. Zhang, W. Wang, Y.-L. Wang, X. Ruan, W. Liu, L. He, R. Zhang, H. Zhang, B. Liu, R. Liu, H. Meng, and Y. Xu, Strong interface-induced spin-charge conversion in YIG/Cr heterostructures, Appl. Phys. Lett. **117**, 112402, (2020).

[32] Q. Chen, W. Lv, S. Li, W. Lv, J. Cai, Y. Zhu, J. Wang, R. Li, B. Zhang, and Z. Zeng, Spin orbit torques in Pt-based heterostructures with van der Waals interface*, Chin. Phys. B **30**, 097506, (2021).

[33] W. L. Peng, J. Y. Zhang, G. N. Feng, X. L. Xu, C. Yang, Y. L. Jia, and G. H. Yu, Enhancement of spin-orbit torque via interfacial hydrogen and oxygen ion manipulation, Appl. Phys. Lett. **115**, 092402, (2019).

[34] H. L. Wang, J. Finley, P. X. Zhang, J. H. Han, J. T. Hou, and L. Q. Liu, Spin-orbit-torque switching mediated by an antiferromagnetic insulator, Phys. Rev. Appl. **11**, 044070, (2019).

[35] C. Y. Guo, C. H. Wan, M. K. Zhao, C. Fang, T. Y. Ma, X. Wang, Z. R. Yan, W. Q. He, Y. W. Xing, J. F. Feng, and X. F. Han, Switching the perpendicular magnetization of a magnetic insulator by magnon transfer torque, Phys. Rev. B **104**, 094412, (2021).

[36] L. J. Zhu, L. J. Zhu, and R. A. Buhrman, Fully spin-transparent magnetic interfaces enabled by the insertion of a thin paramagnetic NiO layer, Phys. Rev. Lett. **126**, 107204(2021).

[37] Q. Li, M. Yang, C. Klewe, P. Shafer, A. T. N'Diaye, D. Hou, T. Y. Wang, N. Gao, E. Saitoh, C. Hwang, R. J. Hicken, J. Li, E. Arenholz, and Z. Q. Qiu, Coherent AC spin current transmission across an antiferromagnetic CoO insulator, Nat. Commun. **10**, 5265, (2019).

[38] X. Shu, J. Zhou, L. Liu, W. Lin, C. Zhou, S. Chen, Q. Xie, L. Ren, Y. Xiaojiang, H. Yang, and J. Chen, Role of interfacial orbital hybridization in spin-orbit-torque generation in Pt-based heterostructures, Phys. Rev. Appl. **14**, 054056, (2020).

[39] W. Zhang, W. Han, X. Jiang, S.-H. Yang, and S. S. P. Parkin, Role of transparency of platinum–ferromagnet interfaces in determining the intrinsic magnitude of the spin Hall effect, Nat. Phys. **11**, 496, (2015).

[40] H. Hayashi, A. Musha, H. Sakimura, and K. Ando, Spin-orbit torques originating from the bulk and interface in Pt-based structures, Phys. Rev. Res. **3**, 013042, (2021).

[41] C.-F. Pai, Y. Ou, L. H. Vilela-Leao, D. C. Ralph, and R. A. Buhrman, Dependence of the efficiency of spin Hall torque on the transparency of Pt/ferromagnetic layer interfaces, Phys. Rev. B **92**, 064426, (2015).

[42] H. Moriya, A. Musha, and K. Ando, Interfacial spin-orbit torque and spin transparency in Co/Pt bilayer, Appl. Phys. Express **14**, 063001, (2021).

[43] T. D. Skinner, M. Wang, A. T. Hindmarch, A. W. Rushforth, A. C. Irvine, D. Heiss, H. Kurebayashi, and A. J. Ferguson, Spin-orbit torque opposing the Oersted torque in ultrathin Co/Pt bilayers, Appl. Phys. Lett. **104**, 062401 (2014).

[44] S. Dutta, A. Bose, A. A. Tulapurkar, R. A. Buhrman, and D. C. Ralph, Interfacial and bulk spin Hall contributions to fieldlike spin-orbit torque generated by iridium, Phys. Rev. B **103**, 184416, (2021).

[45] G. Sala and P. Gambardella, Giant orbital Hall effect and orbital-to-spin conversion in 3d, 5d, and 4f metallic heterostructures, Phys. Rev. Res. **4**, 033037, (2022).

[46] A. B. Drovosekov, N. M. Kreines, A. O. Savitsky, E. A. Kravtsov, M. V. Ryabukhina, V. V. Proglyado, and V. V. Ustinov, Magnetization and ferromagnetic resonance in a Fe/Gd multilayer: Experiment and modelling, J. Phys.: Condens. Matter **29**, 115802, (2017).

[47] A. B. Drovosekov, A. O. Savitsky, D. I. Kholin, N. M. Kreines, V. V. Proglyado, M. V. Makarova, E. A. Kravtsov, and V. V. Ustinov, Twisted magnetization states and inhomogeneous resonance modes in a Fe/Gd ferrimagnetic multilayer, J. Magn. Magn. Mater. **475**, 668, (2019).

[48] K. Zhou, X. Zhan, Z. Li, H. Li, C. Yan, L. Chen, and R. Liu, Efficient characteristics of exchange coupling and spin–flop transition in Py/Gd bilayer using anisotropic magnetoresistance, Appl. Phys. Lett. **122**, 102403, (2023).

[49] L. Zhu, D. C. Ralph, and R. A. Buhrman, Effective spin-mixing conductance of heavy-metal-ferromagnet interfaces, Phys. Rev. Lett. **123**, 057203, (2019).

[50] L. Zhu, L. Zhu, D. C. Ralph, and R. A. Buhrman, Origin of strong two-magnon scattering in heavy-metal/ferromagnet/oxide heterostructures, Phys. Rev. Appl. **13**, 034038, (2020).

[51] M. Zwierzycki, Y. Tserkovnyak, P. J. Kelly, A. Brataas, and G. E. W. Bauer, First-principles study of magnetization relaxation enhancement and spin transfer in thin magnetic films, Phys. Rev. B **71**, 064420, (2005).

[52] G. Tatara and S. Mizukami, Consistent microscopic analysis of spin pumping effects, Phys. Rev. B **96**, 064423, (2017).

[53] J. M. Shaw, H. T. Nembach, T. J. Silva, and C. T. Boone, Precise determination of the spectroscopic g-factor by use of broadband ferromagnetic resonance spectroscopy, J. Appl. Phys. **114**, 243906 (2013).

[54] D. Lee, D. Go, H. J. Park, W. Jeong, H. W. Ko, D. Yun, D. Jo, S. Lee, G. Go, J. H. Oh, K. J. Kim, B. G. Park, B. C. Min, H. C. Koo, H. W. Lee, O. Lee, and K. J. Lee, Orbital torque in magnetic bilayers, Nat. Commun. **12**, 6710, (2021).

[55] S. Dutta and A. A. Tulapurkar, Observation of nonlocal orbital transport and sign reversal of dampinglike torque in Nb/Ni and Ta/Ni bilayers, Phys. Rev. B **106**, 184406, (2022).

[56] S. Lee, M.-G. Kang, D. Go, D. Kim, J.-H. Kang, T. Lee, G.-H. Lee, J. Kang, N. J. Lee, Y. Mokrousov, S. Kim, K.-J. Kim, K.-J. Lee, and B.-G. Park, Efficient conversion of orbital Hall current to spin current for spin-orbit torque switching, Commun. Phys. **4**, 234, (2021).

[57] S. Ding, A. Ross, D. Go, L. Baldrati, Z. Ren, F. Freimuth, S. Becker, F. Kammerbauer, J. Yang, G. Jakob, Y. Mokrousov, and M. Kläui, Harnessing orbital-to-spin conversion of interfacial orbital currents for efficient spin-orbit torques, Phys. Rev. Lett. **125**, 177201, (2020).

[58] S. Ding, Z. Liang, D. Go, C. Yun, M. Xue, Z. Liu, S. Becker, W. Yang, H. Du, C. Wang, Y. Yang,

G. Jakob, M. Kläui, Y. Mokrousov, and J. Yang, Observation of the orbital Rashba-Edelstein magnetoresistance, Phys. Rev. Lett. **128**, 067201, (2022).

[59] V. P. Amin, J. Zemen, and M. D. Stiles, Interface-generated spin currents, Phys. Rev. Lett. **121**, 136805 (2018).

[60] G. Bihlmayer, P. Noël, D. V. Vyalikh, E. V. Chulkov and A. Manchon, Rashba-like physics in condensed matter, Nature Reviews Physics **4**, 642–659 (2022).